\documentclass[journal,10pt]{IEEEtran}

\usepackage{color}
\usepackage{multirow}
\usepackage{graphicx}
\usepackage{epstopdf}
\usepackage[cmex10]{amsmath}
\usepackage{amssymb}

\usepackage{multirow}
\usepackage{amsthm}
\usepackage{cite}
\usepackage{flushend}
\usepackage{acronym} %This package will write out the acronym the first time it is used and write it as an acronym for the rest of the paper
\usepackage{algorithm, algcompatible}

\algnewcommand\INPUT{\item[\textbf{Input:}]}
\algnewcommand\OUTPUT{\item[\textbf{Output:}]}

\allowdisplaybreaks

\begin{document}

\title{Generalized Energy Detection Under Generalized Noise Channels} 
\author{Nikolaos~I.~Miridakis,~\IEEEmembership{Senior Member,~IEEE}, Theodoros~A.~Tsiftsis,~\IEEEmembership{Senior Member,~IEEE} \\and Guanghua~Yang,~\IEEEmembership{Senior Member,~IEEE}
%\thanks{Copyright \copyright  2020 IEEE. Personal use of this material is permitted. However, permission to use this material for any other purposes must be obtained from the IEEE by sending a request to pubs-permissions@ieee.org.}
%\thanks{\textit{Corresponding Author: T. A. Tsiftsis.}}
\thanks{N. I. Miridakis, T. A. Tsiftsis and G. Yang are with the Institute of Physical Internet and School of Intelligent Systems Science \& Engineering, Jinan University, Zhuhai Campus, Zhuhai 519070, China. N. I. Miridakis is also with the Dept. of Informatics and Computer Engineering, University of West Attica, Aegaleo 12243, Greece (e-mails: nikozm@uniwa.gr, theo\_tsiftsis@jnu.edu.cn, ghyang@jnu.edu.cn).}
}

%\markboth{}{}

\maketitle

\begin{abstract}
Generalized energy detection (GED) is analytically studied when operates under fast-faded channels and in the presence of generalized noise. For the first time, the McLeish distribution is used to model the underlying noise, which is suitable for both non-Gaussian (impulsive) as well as classical Gaussian noise channels. Important performance metrics are presented in closed forms, such as the false-alarm and detection probabilities as well as the decision threshold. Analytical and simulation results are cross-compared validating the accuracy of the proposed approach in the entire signal-to-noise ratio regime. Finally, useful outcomes are extracted with respect to GED system settings under versatile noise environments and when noise uncertainty is present.
\end{abstract}


\IEEEpeerreviewmaketitle

\section{Introduction}
\IEEEPARstart{S}{ignal} detection and spectrum sensing have attracted a vast research interest over the last decades, whereas they are considered as essential counterparts of various practical applications. The most representative ones include cognitive radio (CR) transmissions, radar communications as well as network slicing and dynamic frequency resource allocation in 5G networks \cite{j:FoukasPatounas17,j:JiangLiYe2019,j:YeLuLi2017}. Among the available signal detection schemes, energy detection (ED) is one of the most popular ones because it provides an efficient tradeoff between computational complexity and performance \cite{j:Urkowitz67}. In fact, ED is a relatively simple implementation approach while it is the optimum detector in the presence of Gaussian signals and additive white Gaussian noise (AWGN) channels \cite{j:Yucek2009}. Further, it does not require any knowledge regarding the signal and channel fading statistics.

Nonetheless, in realistic conditions, neither the transmitted signals are always Gaussian distributed nor the underlying noise is AWGN. Particularly, there are various types of wireless communication channels where signals are subjected to non-Gaussian (i.e., impulsive with heavy-tailed distributional behavior) noise. Typical examples include urban and indoor wireless channels, ultra-wide band communications, frequency/time-hopping with jamming, millimeter wave communications, and wireless transmissions under strong interference conditions (e.g., see \cite{j:MoghimiNasriSchober11,j:ZhuWang2019} and relevant references therein). Accordingly, a modified type of ED has been thoroughly analyzed and tested, which is entitled as generalized ED (GED) or $p$-norm detector, where $p$ is a tunable parameter so as to enhance the detector performance. GED includes as special cases the absolute value detector when $p=1$ \cite{j:GaoLi2015}, ED when $p=2$ and fractional low order detector when $0<p<2$ \cite{j:MoghimiNasriSchober11,j:ZhuWang2019}. The performance of GED was studied in \cite{j:YeLi2019} for a certain popular type of non-Gaussian impulsive noise channels; namely, additive white Laplacian noise (AWLN). Also, GED was studied in the presence of generalized noise in \cite{j:KostylevGres18}, by using the Gaussian mixture distribution model. However, the derived results were quite complex, i.e., defined in an infinite series representation. On a similar basis, \cite{j:MoghimiNasriSchober11} and \cite{j:ZhuWang2019} studied GED under non-Gaussian noise channels; yet, their detection performance results were tightly accurate in the case when the noise power is much higher than the signal power.

In this Letter, for the first time, we analytically study the GED performance under McLeish noise channels. McLeish distribution represents a generalized model, appropriate for both Gaussian and non-Gaussian noise channels. It was originated by D. Mcleish in \cite{j:McLeish82} and quite recently it was revisited and thoroughly analyzed in \cite{j:YilmazMcLeish2020}. McLeish distribution resembles the Gaussian distribution; it is unimodal, symmetric, it has all its moments finite, and has tails that are at least as heavy as those of Gaussian distribution. Moreover, the evolution of its impulsive nature from Gaussian distribution to non-Gaussian distribution is explicitly parameterized in a rigorous way with psychical meaning (please, see the detailed analysis in \cite[\S IV.B.]{j:YilmazMcLeish2020}); especially than those of Laplacian, $\alpha$-stable and generalized Gaussian distributions. It models any kind of impulsive noise between the two extreme cases, i.e., Dirac's distribution (highly impulsive noise) and AWGN (non-impulsive noise). 

For sufficiently large number of samples, which is usually the practical case, analytical closed-form expressions are derived for key performance metrics, namely, the false-alarm and detection probabilities as well as the decision threshold of GED. The mentioned expressions are sharp in the entire signal-to-noise (SNR) ratio regime. In addition, the case when the received signal undergoes fast-faded Rician channels is included as well as the detrimental effect of uncertain/imperfect noise power estimation. Finally, the enclosed analytical and numerical results reveal some useful engineering insights.

{\it Notation:} $|\cdot|$ represents absolute (scalar) value. $\mathbb{E}[\cdot]$ is the expectation operator, ${\rm Var}[\cdot]$ is the variance operator and symbol $\overset{\text{d}}=$ means equality in distribution. $f_{x}(\cdot)$ denotes the probability density function (PDF) of random variable (RV) $x$. Also, $y_{|z}$ denotes that $y$ is conditioned on $z$ event. $\mathcal{CN}(\mu,\sigma^{2})$ and $\mathcal{N}(\mu,\sigma^{2})$ define, respectively, a complex and circularly symmetric (CCS) Gaussian RV as well as a real-valued Gaussian RV with mean $\mu$ and variance $\sigma^{2}$. Moreover, $\mathcal{CML}(\mu,\sigma^{2},v)$ denotes a CCS RV following the McLeish distribution with mean $\mu$, variance $\sigma^{2}$ and non-Gaussianity parameter $v$. Further, $Q(\cdot)$ and $Q^{-1}(\cdot)$ are the Gaussian $Q$-function and inverse $Q$-function, respectively, while ${\rm csc(\cdot)}$ stands for the cosecant function. $\Gamma(\cdot)$ denotes the Gamma function \cite[Eq. (8.310.1)]{tables} and $\Gamma(\cdot,\cdot)$ is the upper incomplete Gamma function \cite[Eq. (8.350.2)]{tables}. $I_{0}(\cdot)$ is the $0^{\rm th}$ order modified Bessel function of the first kind \cite[Eq. (8.445)]{tables}; $K_{v}(\cdot)$ denotes the $v^{\rm th}$ order modified Bessel function of the second kind \cite[Eq. (8.432)]{tables}; ${}_1F_{1}(\cdot,\cdot;\cdot)$ is the Kummer confluent hypergeometric function \cite[Eq. (9.210.1)]{tables}; and $G^{m,n}_{p,q}[\cdot|\cdot]$ represents the Meijer's G-function \cite[Eq. (9.301)]{tables}. Finally, ${\rm Re}\{x\}$ and ${\rm Im}\{x\}$ denote the real and imaginary part of a complex-valued $x$, respectively.

\section{System and Signal Model}
Consider the binary hypothesis problem, which reads as
\begin{align}
\begin{array}{l l l}     
    \mathcal{H}_{0}: &y[u]=w[u],& \text{no signal is present,} \\
    \mathcal{H}_{1}: &y[u]=h[u] s[u]+w[u],  & \text{signal transmission,}
\end{array}
\label{proform}
\end{align}
where $y[u] \in \mathbb{C}$, $h[u]\in \mathbb{C}$, $s[u] \in \mathbb{R}$ and $w[u] \in \mathbb{C}$ denote the received signal, channel fading coefficient, transmitted baseband signal and additive noise, respectively, at the $u^{\rm th}$ sample. The signal samples, $s[\cdot]$, are being transmitted with power $s^{2}$, and are subjected to an arbitrary continuous or discrete distribution. Further, it is assumed that the channel fading coefficient, $h[\cdot]$, follows a non zero mean CCS Gaussian distribution, such that ${\rm Re}\{h[\cdot]\}\overset{\text{d}}=\mathcal{N}({\rm cos}(\theta)\alpha,\sigma^{2}_{h}/2)$ and ${\rm Im}\{h[\cdot]\}\overset{\text{d}}=\mathcal{N}({\rm sin}(\theta)\alpha,\sigma^{2}_{h}/2)$ for arbitrary $\theta \in [0,2\pi)$ and $\alpha \in \mathbb{R}^{+}$, while $\sigma^{2}_{h}$ stands for the variance of $h[\cdot]$. Consequently, $|h[\cdot]|$ follows the Rice distribution with Rician factor $K\triangleq \alpha^{2}/\sigma^{2}_{h}$, which sufficiently models both line-of-sight (LoS) and non-LoS channel fading conditions; note that $|h[\cdot]|$ becomes Rayleigh distributed for $\alpha=K=0$. It is also assumed that $h[\cdot]$ remains fixed during a sample time whereas it may change between different samples.

In addition, $w[\cdot]\overset{\text{d}}=\mathcal{CML}(0,\sigma^{2}_{w},v)$ with $\sigma^{2}_{w}\in \mathbb{R}^{+}$ and $v\in \mathbb{R}^{+}$ standing for the noise variance and non-Gaussianity parameter, respectively, with a symmetric and unimodal PDF defined as \cite[Eq. (85)]{j:YilmazMcLeish2020}
\begin{align}
f_{w}(w)=\frac{2 \sqrt{v} |w|^{v-1}}{\sqrt{2 \sigma^{2}_{w}} \pi \Gamma(v)}K_{v-1}\left(\sqrt{\frac{2 v}{\sigma^{2}_{w}}} |w|\right).
\label{noisePDF}
\end{align}
Some special cases of $f_{w}(\cdot)$ are obtained for $v=1$, $v\rightarrow +\infty$ and $v\rightarrow 0^{+}$ resulting to the CCS Laplacian, Gaussian and Dirac's distribution, respectively \cite{j:YilmazMcLeish2020}. It turns out that the McLeish distribution is a generalized and versatile distribution model, which is suitable for both Gaussian and non-Gaussian (impulsive) noise channels. 

Moreover, GED is fully unaware of channel gains as well as the signal and noise statistics; reflecting on a \emph{blind} spectrum sensing. The considered test statistic for GED reads as
\begin{align}
T\triangleq \sum^{N}_{u=1}|y[u]|^{p},
\label{teststat}
\end{align}
where $N$ represents the number of samples and $p\geq 0$ is a tunable exponent that provides flexibility to the detector. When $p=2$, GED coincides with the conventional ED, while it becomes the fractional low order detector for $0<p<2$.

\section{Performance Metrics}
The scenario of a false-alarm probability, namely, $P_{f}(\cdot)$, is modeled by $P_{f}(\lambda)\triangleq \text{Pr}[T>\lambda|\mathcal{H}_{0}]$ with $\lambda$ denoting the decision threshold. For sufficiently large number of samples, which is usually the practical case, the PDF of $T$ closely approaches a Gaussian distribution even if the underlying noise (having finite moments) is non-Gaussian. Thus, for arbitrary $p$, the false-alarm probability is presented in a simple closed form as
\begin{align}
P_{f}(\lambda)&=Q\left(\frac{\lambda-N \mu_{0}}{\sqrt{N \sigma^{2}_{0}}}\right),
\label{Pf}
\end{align}
where\footnote{Since consecutive samples are mutually independent and for notational simplicity, hereinafter we drop sample indexing.}
\begin{align}
\mu_{0}\triangleq \mathbb{E}[|y|^{p}_{|\mathcal{H}_{0}}]=\mathbb{E}[|w|^{p}],
\label{m0}
\end{align}
and
\begin{align}
\sigma^{2}_{0}\triangleq {\rm Var}[|y|^{p}_{|\mathcal{H}_{0}}]=\mathbb{E}[|w|^{2 p}]-\mathbb{E}[|w|^{p}]^{2},
\label{s0}
\end{align}
with
\begin{align}
\mathbb{E}[|w|^{n}_{|\mathcal{H}_{0}}]=\frac{\Gamma\left(\frac{n}{2}+v\right)\Gamma\left(\frac{n}{2}+1\right)}{\Gamma\left(v\right)v^{n/2}}\sigma^{n}_{w},\quad n \in \mathbb{R}.
\label{mom0}
\end{align}
The proof of \eqref{Pf} is relegated in Appendix \ref{appa}.

As it is obvious from (\ref{Pf}), the false-alarm probability is an \emph{offline} operation, i.e., it is independent of channel gains and signal statistics. For known $N$ and $\sigma^{2}_{w}$, the common practice of setting the decision threshold is based on the constant false-alarm probability. Also, this is a reasonable assumption since for various practical spectrum sensing applications, the highest priority is to satisfy a predetermined false-alarm rate (e.g., underlay CR). Doing so, the desired threshold, $\lambda^{\star}$, stems as
\begin{align}
\lambda^{\star}\triangleq Q^{-1}\left(P^{({\tau})}_{f}\right)\sqrt{N \sigma^{2}_{0}}+N \mu_{0},
\label{thr}
\end{align} 
where $P^{({\tau})}_{f}$ represents the predetermined target on the maximum attainable false-alarm probability.

In the case of signal transmission, modeled by the $\mathcal{H}_{1}$ hypothesis, the detection probability, $P_{d}(\cdot)$, is directly obtained in a closed form as
\begin{align}
\nonumber
P_{d}(\lambda^{\star})&\triangleq \text{Pr}[T>\lambda|\mathcal{H}_{1}]\\
&=Q\left(\frac{\lambda^{\star}-N \mu_{1}}{\sqrt{N \sigma^{2}_{1}}}\right),
\label{Pd}
\end{align} 
where 
\begin{align}
\mu_{1}\triangleq \mathbb{E}[|y|^{p}_{|\mathcal{H}_{1}}]=\mathbb{E}[|h s+w|^{p}],
\label{m1}
\end{align}
and
\begin{align}
\sigma^{2}_{1}\triangleq {\rm Var}[|y|^{p}_{|\mathcal{H}_{1}}]=\mathbb{E}[|h s+w|^{2 p}]-\mathbb{E}[|h s+w|^{p}]^{2},
\label{s1}
\end{align}
with
\begin{align}
\nonumber
\mathbb{E}[|h s+w|^{n}_{|\mathcal{H}_{1}}]&=\frac{\Gamma\left(1+\frac{n}{2}\right)\left(\frac{s^{2}\sigma^{2}_{h}}{1+\alpha^{2}}\right)^{n/2}}{\Gamma\left(v\right) \Gamma\left(-\frac{n}{2}\right)}{}_1F_{1}\left(-\frac{n}{2},1;-\alpha^{2}\right)\\
&\times G^{1,2}_{2,1}\left[\frac{\sigma^{2}_{w}}{s^{2}\sigma^{2}_{h} v}~\vline
\begin{array}{c}
1-v,\frac{n}{2}+1 \\
0 
\end{array} \right],
\label{mom1}
\end{align}
for arbitrary $n$ excluding any even integer $n\geq 2$. For the latter case (where $n=2,4,6,\ldots$), \eqref{mom1} relaxes to 
 \begin{align}
\nonumber
\mathbb{E}[|h s+w|^{n}_{|\mathcal{H}_{1}}]&=\frac{\Gamma\left(1+\frac{n}{2}\right)}{\Gamma(v)(1+\alpha^{2})^{n/2}}{}_1F_{1}\left(-\frac{n}{2},1;-\alpha^{2}\right)\\
&\times \sum^{n/2}_{k=0}\binom{n/2}{k}\left(s^{2}\sigma^{2}_{h}\right)^{n/2-k}\sigma^{2 k}_{w}\Gamma(k+v)v^{-k}.
\label{mom11}
\end{align}
The proof of \eqref{Pd} is provided in Appendix \ref{appb}. Note that for Rayleigh faded channels, ${}_1F_{1}\left(-n/2,1;0\right)=1$ in \eqref{mom1} and \eqref{mom11}. In addition, for the special case of CCS Laplacian noise (i.e., when $v=1$), the Meijer's G-function in \eqref{mom1} reduces to 
\begin{align}
\nonumber
G^{1,2}_{2,1}\left[\frac{\sigma^{2}_{w}}{s^{2}\sigma^{2}_{h}}~\vline
\begin{array}{c}
0,\frac{n}{2}+1 \\
0 
\end{array} \right]&=-\pi \exp\left(-\frac{\sigma^{2}_{w}}{s^{2}\sigma^{2}_{h}}\right) \left(\frac{\sigma^{2}_{w}}{s^{2}\sigma^{2}_{h}}\right)^{n/2}\\
&\times {\rm csc}\left(\frac{\pi n}{2}\right) \frac{\Gamma\left(\frac{n}{2}+1,\frac{s^{2}\sigma^{2}_{h}}{\sigma^{2}_{w}}\right)}{\Gamma\left(\frac{n}{2}+1\right)}.
\label{GfunctionReduction}
\end{align}

In the presence of detrimental yet unavoidable effect of uncertain noise power estimation, the corresponding uncertainty factor can be modeled such that $\sigma^{2}_{w}\in \{\hat{\sigma}^{2}_{w}/\rho,\rho \hat{\sigma}^{2}_{w}\}$ with $\hat{\sigma}^{2}_{w}$ standing for the estimated noise power and $\rho\geq 1$ \cite{j:Tandra}. In practice, the distribution of the actual uncertainty factor is quite difficult to obtain. However, the bound on the mentioned uncertainty (i.e., $\rho$) is measurable\footnote{As an illustrative example, IEEE 802.22 and ECMA 392 standards utilize sporadic long sensing periods for fine sensing and more frequent short sensing periods in which a variety of signal-free samples can be collected and further processed for noise estimation \cite{j:SenanayakeSmith20}.} and thus can be considered as known. To evaluate the worst-case scenario (i.e., the lower bound on $P_{d}(\cdot)$ given a fixed $P^{({\tau})}_{f}$), \eqref{mom0} and \eqref{mom1} are directly computed by substituting $\sigma^{2}_{w}$ with $\rho \hat{\sigma}^{2}_{w}$. Doing so and according to \eqref{thr}, it turns out that $\lambda^{\star}(\rho)\triangleq \rho^{p/2} \lambda^{\star}$, reflecting on a corresponding deviation on the decision threshold (thus, on the detector performance) which is proportional to $p/2$. Hence, it is verified that detectors based on fractional low order statistics are indeed more robust to noise uncertainty. Yet, setting $p\rightarrow 0^{+}$ is not always effective since \eqref{Pd} is a non-concave function with respect to the exponent $p$ or the non-Gaussianity parameter $v$. Hence, to obtain the optimum $p^{\star}$, it is required to solve
\begin{align}
\underset{p}{{\rm arg}\max}\:P_{d}(\lambda^{\star}),\quad \textrm{subject to } \{v,p\}>0,
\end{align}
which can be numerically computed by a simple line search via \eqref{thr} and \eqref{Pd}.   

\section{Numerical Results and Discussion}
\begin{figure}[!t]
\centering
\includegraphics[trim=1.5cm 0.2cm 0.5cm .1cm, clip=true,totalheight=0.25\textheight]{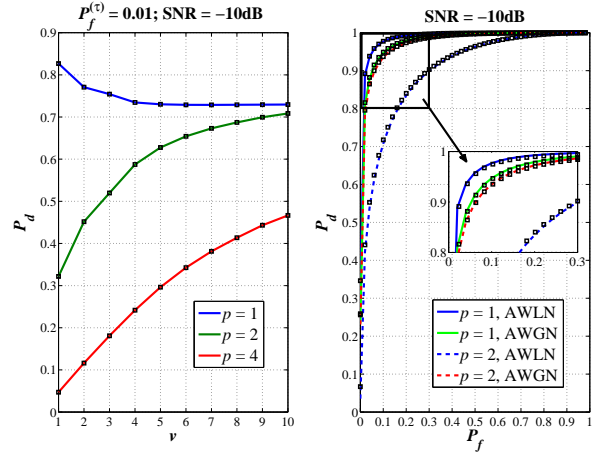}
\caption{GED performance for various system parameters and noise channels.}
\label{fig1}
\end{figure}

\begin{figure}[!t]
\centering
\includegraphics[trim=1.5cm 0.2cm 0.5cm .1cm, clip=true,totalheight=0.25\textheight]{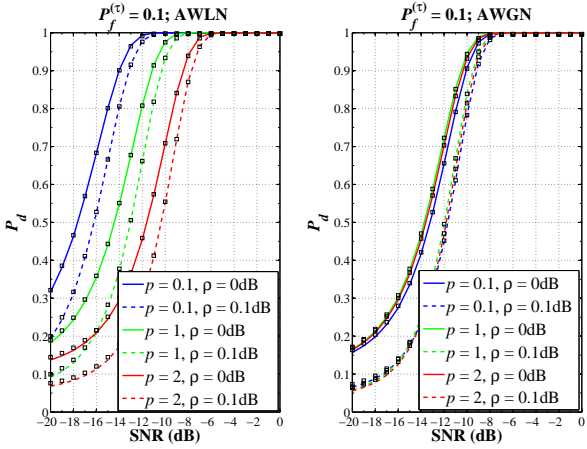}
\caption{Detection probability of GED vs. various SNR regions for different noise channels and system settings.}
\label{fig2}
\end{figure}

In this section, the derived analytical results are verified via numerical validation, whereas they are cross-compared with corresponding Monte-Carlo simulations. Subsequently, in Figs.~\ref{fig1} and \ref{fig2}, $h[u]\overset{\text{d}}=\mathcal{CN}(0,1)$ for the $u^{\rm th}$ sample; reflecting on unit-scale Rayleigh fast-faded channels. In Fig.~\ref{fig3}, $h[u]\overset{\text{d}}=\mathcal{CN}(\alpha,1)$ denoting Rician channel fading with a corresponding factor $K=\alpha^{2}$. Hence, the received SNR is defined as $\text{SNR}\triangleq s^{2}/\sigma^{2}_{w}$. All the simulation results are conducted by averaging $10^{4}$ independent trials. Hereinafter, line-curves and square-marks denote the analytical and simulation results, respectively, while the number of samples is set to be $N=2^{10}$.  

At the left-hand side (LHS) of Fig.~\ref{fig1}, the impact of non-Gaussianity parameter $v$ on the detection probability is depicted. When the noise becomes more impulsive (i.e., a reduced $v$), a lower $p$ exponent is beneficial. In fact, this effect gets even more emphatic as $v\rightarrow 0^{+}$. On the other hand, as the noise tends to approach the Gaussian type (i.e., for an increased $v$), the detection performance of a low-order $p$ is degraded. At the right-hand side (RHS) of Fig.~\ref{fig1}, the receiver operating characteristic curve of GED is shown for two certain noise types, namely, AWLN ($v=1$) and AWGN ($v\rightarrow +\infty$), while comparing the absolute value detector ($p=1$) and ED ($p=2$). Obviously, the former detector outperforms the latter one in the case of Laplacian noise (highly impulsive noise), whereas quite the opposite outcome arises from the Gaussian noise scenario.

In Fig.~\ref{fig2}, the GED performance is illustrated for the practical case of noise uncertainty. Particularly, the worst-case scenario of the detection performance is shown for a given/known uncertainty bound $\rho$ (perfect noise power estimation is defined as $\rho=0$dB). The LHS and RHS of Fig.~\ref{fig2} correspond to the AWLN and AWGN, respectively. A number of useful engineering insights can be drawn from Fig.~\ref{fig2}. Reducing the $p$ exponent of GED in AWLN channels enhances the detection performance both in the presence and absence of noise uncertainty. Indicatively, setting $p=0.1$ provides approximately a $6$dB gain against the classical ED ($p=2$). Nevertheless, an entirely different behavior is observed in AWGN channels, where ED presents a better detection performance whilst the corresponding performance of the low-order detector with $p=0.1$ degrades. It is also noteworthy that the detection performance of all the considered GED cases almost coincide to each other in the case of imperfect noise estimation; thus revealing the detrimental effect of noise uncertainty regardless of the value of $p$ exponent in Gaussian noise. 

{\color{black}In Fig.~\ref{fig3}, the case of Rician faded channels is illustrated for two additive noise models; namely, AWLN and AWGN. The practical scenario of noise uncertainty is also included (i.e., $\rho=0.1$dB) as well as the ideal case of perfect noise estimation ($\rho=0$dB). Moreover, $K=0$ denotes the classical Rayleigh fading, whereas $K=10$ indicates the presence of a dominant (LoS) factor. The detection performance is being enhanced for the ideal noise estimation case and when the received signal undergoes Rician fading. This is a reasonable outcome since the presence of a strong dominant signal power factor makes the actual received signal more distinguishable than additive noise. From an engineering standpoint, the non-concavity of the detection performance with respect to exponent $p$ is evident under various types of channel fading and/or noise environments. Obviously, under Rayleigh faded channels, energy-type detectors (i.e., $p\propto 2$) are suitable for both ideal and non-ideal noise estimation. Nevertheless, fractional low order detectors are much more beneficial for an increased Rician $K$ factor or when the noise becomes impulsive, since the detection performance is being enhanced as $p$ is reduced. Finally, it is worthy to state that an integer-valued exponent $p$ produces less computational complexity than its fractional-order counterpart \cite{j:YeLi2019}. Thereby, whenever the complexity reduction is of prime importance, absolute value detector with $p=1$ represents quite an effective detector (as compared to ED or higher order detectors) in the presence of Rician channel fading, impulsive noise and/or noise uncertainty.

\begin{figure}[!t]
\centering
\includegraphics[trim=0.5cm 0.2cm 0.5cm .1cm, clip=true,totalheight=0.25\textheight]{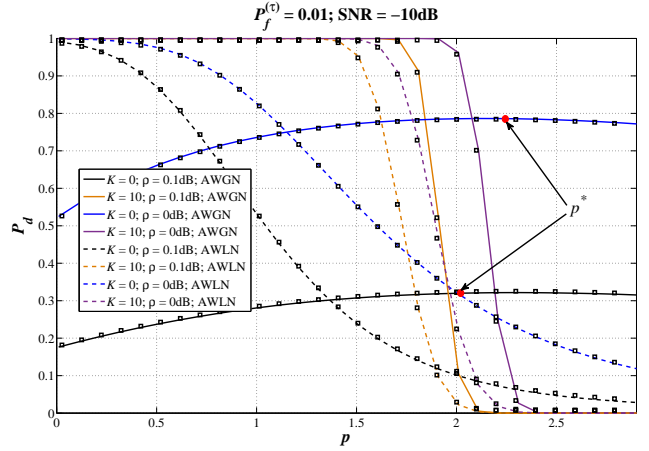}
\caption{Detection probability of GED vs. various exponent $p$ values for different fading channels and noise conditions.}
\label{fig3}
\end{figure}

\section{Conclusion}
The GED (also known as $p$-norm detector) was analytically studied under the presence of Rician faded channels and McLeish noise, thus capturing a wide range from highly impulsive to non-impulsive noise conditions. Important system performance metrics were derived in straightforward closed-form expressions; namely, the decision threshold, detection and false-alarm probabilities. Capitalizing on these expressions, the optimum exponent of GED can be numerically computed quite easily. Finally, some useful outcomes have been manifested including the case when the detrimental yet unavoidable effect of noise uncertainty is present.  
}

\appendix
\subsection{Derivation of the $n^{\textrm{th}}$-moment function for hypothesis $\mathcal{H}_{0}$}
\label{appa}
\numberwithin{equation}{subsection}
\setcounter{equation}{0}
We commence by decomposing $w$ to the product of the squared-root of a Gamma distributed RV and a CCS Gaussian RV, which are mutually independent \cite[Thm. 10]{j:YilmazMcLeish2020}, i.e., $w=\sqrt{G} X$, where $X\overset{\text{d}}=\mathcal{CN}(0,\sigma^{2}_{w})$ and
\begin{align}
f_{G}(g)=\frac{v^{v}}{\Gamma(v)}g^{v-1}\exp(-v g), \quad g \in \mathbb{R}^{+}.
\label{pdfg}
\end{align}
Then, $\mathbb{E}[|w|^{n}]=\mathbb{E}[G^{n/2}]\mathbb{E}[|X|^{n}]$. Since $|X|$ is Rayleigh distributed, it is straightforward to show that
\begin{align}
\mathbb{E}[|X|^{n}]=\Gamma\left(\frac{n}{2}+1\right) \sigma^{n}_{w}. 
\label{rayleighmom}
\end{align}
Further, it holds that 
\begin{align}
\mathbb{E}[G^{\frac{n}{2}}]=\frac{v^{v}}{\Gamma(v)}\int^{+\infty}_{0}g^{n/2+v-1}\exp(-v g)dg=\frac{\Gamma\left(\frac{n}{2}+v\right)}{\Gamma\left(v\right) v^{\frac{n}{2}}}.
\label{gammamom}
\end{align}
Combining \eqref{rayleighmom} and \eqref{gammamom}, we arrive at \eqref{mom0}.

\subsection{Derivation of the $n^{\textrm{th}}$-moment function for hypothesis $\mathcal{H}_{1}$}
\label{appb}
\numberwithin{equation}{subsection}
\setcounter{equation}{0}
Following the same lines of reasoning as in Appendix \ref{appa}, we get
\begin{align}
y=h s+w=\underbrace{h s}_{\triangleq z_{1}} +\underbrace{\sqrt{G} X}_{\triangleq z_{2}},
\label{zdefinition}
\end{align}
where $G$ and $X$ are defined in Appendix \ref{appa}. Conditioned on $s$ and $G$ and utilizing the linear properties of Gaussian RVs, it holds that $z_{1}\overset{\text{d}}=\mathcal{CN}(\alpha,s^{2} \sigma^{2}_{h})$ and $z_{2}\overset{\text{d}}=\mathcal{CN}(0,G \sigma^{2}_{w})$. Recall that $h,s,G$ and $X$ are all mutually independent RVs. Thereby, by introducing the auxiliary variable $r\triangleq z_{1}+z_{2}$, we have that
\begin{align}
\nonumber
r_{|s,G}&\overset{\text{d}}=\mathcal{CN}(\alpha,s^{2} \sigma^{2}_{h}+G \sigma^{2}_{w})\\
&=(s^{2} \sigma^{2}_{h}+G \sigma^{2}_{w})^{\frac{1}{2}}\times \underbrace{\mathcal{CN}(\alpha,1)}_{\triangleq z_{0}}.
\label{zzdefinition}
\end{align}
Thus, the absolute moments of $r$ are given by
\begin{align}
\mathbb{E}[|r|^{n}]=\mathbb{E}\left[(s^{2} \sigma^{2}_{h}+G \sigma^{2}_{w})^{\frac{n}{2}}\right] \mathbb{E}\left[|z_{0}|^{n}\right].
\label{absmomr}
\end{align}

It follows that $|z_{0}|$ is Rice distributed with PDF \cite[Eq. (2.17)]{b:Simon}
\begin{align}
\nonumber
f_{|z_{0}|}(z)=&2 (1+\alpha^{2}) z \exp(-(1+\alpha^{2}) z^{2}-\alpha^{2})\\
&\times I_{0}(2 \alpha z \sqrt{1+\alpha^{2}}),\quad z\geq 0,
\label{pdfz0}
\end{align}
and, hence, with the aid of \cite[Eq. (2.15.5.4)]{PrudnikovVol2}, we get
\begin{align}
\mathbb{E}\left[|z_{0}|^{n}\right]=\frac{\Gamma\left(\frac{n}{2}+1\right)}{(1+\alpha^{2})^{n/2}}{}_1F_{1}\left(-\frac{n}{2},1;-\alpha^{2}\right).
\label{momz00}
\end{align}

Regarding the remaining factor of \eqref{absmomr}, it is required to average out the Gamma distributed $G$ parameter. We first utilize the following transformation \cite[Eq. (8.4.2.5)]{PrudnikovVol3}
\begin{align}
(s^{2} \sigma^{2}_{h}+G \sigma^{2}_{w})^{\frac{n}{2}}=\frac{(s^{2} \sigma^{2}_{h})^{\frac{n}{2}}}{\Gamma\left(-\frac{n}{2}\right)} G^{1,1}_{1,1}\left[\frac{\sigma^{2}_{w} G}{s^{2}\sigma^{2}_{h}}~\vline
\begin{array}{c}
\frac{n}{2}+1 \\
0 
\end{array} \right],
\label{trans}
\end{align}
which is valid for any $n$ except those that are positive even numbers greater or equal to $2$. This is due to the fact that, whenever $n=2,4,6,\ldots$, the Gamma function in the denominator of \eqref{trans} returns singularity. Then, using \eqref{pdfg}, \eqref{trans} and utilizing \cite[Eq. (2.24.3.1)]{PrudnikovVol3}, it yields
\begin{align}
\nonumber
\mathbb{E}\left[(s^{2} \sigma^{2}_{h}+G \sigma^{2}_{w})^{\frac{n}{2}}\right]=&\frac{(s^{2} \sigma^{2}_{h})^{\frac{n}{2}}}{\Gamma(v) \Gamma\left(-\frac{n}{2}\right)}\\
&\times G^{1,2}_{2,1}\left[\frac{\sigma^{2}_{w}}{s^{2}\sigma^{2}_{h} v}~\vline
\begin{array}{c}
1-v,\frac{n}{2}+1 \\
0 
\end{array} \right].
\label{transmom}
\end{align}
Therefore, combining \eqref{momz00} and \eqref{transmom}, we reach \eqref{mom1}. For the alternative case where $n\geq 2$ is an even integer, while using the binomial expansion in the left-hand side of \eqref{trans}, we arrive at \eqref{mom11} after some straightforward manipulations.

\bibliographystyle{IEEEtran}
\bibliography{IEEEabrv,References}

\vfill

\end{document}